\documentclass[aps,amsmath,onecolumn, showpacs,superscriptaddress,pre,10.5pt,nofootinbib]{revtex4-1}
\usepackage{amssymb}
\usepackage{amsmath}
\usepackage{graphicx}
\usepackage{array}
\usepackage{physics}
\usepackage{mathtools}
\usepackage{amssymb}
\usepackage{verbatim}
\usepackage{color}
\usepackage{bm}
\usepackage{bbm}
\usepackage{amsthm}
\usepackage[normalem]{ulem}

\usepackage{color}
\usepackage[dvipsnames]{xcolor}
\usepackage[latin1]{inputenc}
\definecolor{Blue}{rgb}{0.00, 0.00, 1.00}
\definecolor{Red}{rgb}{1.00, 0.00, 0.00}
\definecolor{Green}{rgb}{0.00, 0.60, 0.00}

\usepackage[pdfpagemode=TseNone,
		bookmarksopen=false,
		colorlinks=true,
		urlcolor=blue,
		citecolor=red,
		citebordercolor=blue,
		linkcolor=blue, 
		urlcolor=black, 
		filecolor=red]{hyperref}
		
\newcommand{\bea}{\begin{eqnarray}}
\newcommand{\eea}{\end{eqnarray}}
\newcommand{\be}{\begin{equation}}
\newcommand{\ee}{\end{equation}}
\newcommand{\nn}{\nonumber}
\newcommand{\bee}{\begin{equation*}}
\newcommand{\eee}{\end{equation*}}

\colorlet{Mycolor1}{green!10!orange!90!}

\def\XXint#1#2#3{{\setbox0=\hbox{$#1{#2#3}{\int}$}
     \vcenter{\hbox{$#2#3$}}\kern-.5\wd0}}

\begin{document}

\title{Constrained non-crossing Brownian motions, fermions and the Ferrari-Spohn distribution}

\author{Tristan Gauti\'e}
\affiliation{Laboratoire de Physique de l'Ecole Normale Sup\'erieure,
PSL University, CNRS, Sorbonne Universit\'e, Universit\'e de Paris,
24 rue Lhomond, 75231 Paris, France}
\author{Naftali R. Smith}
\email{naftalismith@gmail.com}
\affiliation{Laboratoire de Physique de l'Ecole Normale Sup\'erieure,
PSL University, CNRS, Sorbonne Universit\'e, Universit\'e de Paris,
24 rue Lhomond, 75231 Paris, France}
\affiliation{LPTMS, CNRS, Universit\'e Paris-Sud, Universit\'e  Paris-Saclay, 91405 Orsay, France} 

\begin{abstract}

A conditioned stochastic process can display a very different behavior from the unconditioned process. In particular, a conditioned process can exhibit non-Gaussian fluctuations even if the unconditioned process is Gaussian. In this work, we revisit the Ferrari-Spohn model of a Brownian bridge conditioned to avoid a moving wall, which pushes the system into a large-deviation regime. We extend this model to an arbitrary number $N$ of non-crossing Brownian bridges. We obtain the joint distribution of the distances of the Brownian particles from the wall at an intermediate time in the form of the determinant of an $N\times N$ matrix whose entries are given in terms of the Airy function. We show that this distribution coincides with that of the positions of $N$ spinless noninteracting fermions trapped by a linear potential with a hard wall. We then explore the $N \gg 1$ behavior of the system. For simplicity we focus on the case where the wall's position is given by a semicircle as a function of time, but we expect our results to be valid for any concave wall function.

\end{abstract}


\maketitle

{%
	\hypersetup{linkcolor=black}
	\tableofcontents
}

\section{Introduction}

An important class of models in probability theory and statistical mechanics is
random processes, conditioned on non-absorption by moving walls.
When this conditioning ``pushes'' the random process into a large-deviation regime, it can lead to rich and interesting behavior; in particular, to non-Gaussian fluctuations, even if the unconditioned process is Gaussian.
Ferrari and Spohn \cite{FerrariSpohn05} considered a Brownian bridge $A(\tau)$ conditioned to stay away from a wall $g(\tau)$. 
They found that typical fluctuations of the distance between the Brownian particle and the wall at the mid-time of the Brownian bridge are described by a universal distribution which is given in terms of the square of the Airy function%
\footnote{
Large deviations in the Ferrari-Spohn (FS) model and in its extensions were studied in \cite{GeomOpt1,GeomOpt2}.}%
.
This distribution has by now been encountered in several other settings: in \cite{Agranov19} for a Brownian excursion conditioned so that the area under it is very small,  
and also in several models for Brownian motion in 2 spatial dimensions conditioned on non-absorption by a stationary wall and pushed into a large-deviation regime through further constraints \cite{Nechaev19,GeomOpt3, Vladimirov20,Valov20}. 
Brownian excursions, conditioned to stay away from a moving wall, have attracted attention in the mathematical literature too, with many extensions and generalizations \cite{Ioffe2015,Ioffe2018a,Ioffe2018b,Ioffe2018c}.

It is natural to consider the extension of models such as those that are described above to an arbitrary number $N$ of (possibly interacting) Brownian particles. 
Indeed, in \cite{NonCrossing}, the problem of non-crossing Brownian motions under a square root barrier $h(t) = W \sqrt{t}$ was studied. It was shown that this problem has deep links with a system of $N$ noninteracting fermions in a harmonic potential with a hard-wall placed at $W$ such that the particles are confined under the wall. The probability for $N$ independent Brownian motions to have not crossed each other or the square root barrier until time $t$ was shown to decay asymptotically as $t^{-\beta(N, W)}$ where $\beta(N, W)$ is the ground state energy of the $N$-fermion system. The joint distribution of the positions of the surviving particles under the barrier was also found explicitly in terms of the quantum propagator of the fermion system. The results from Ref.~\cite{NonCrossing} are a key ingredient for the computation of the present work, see below. In the presence a fixed barrier, the problem of non-crossing Brownian motions described above was studied and linked to fermion systems in \cite{Fisher84,KratGutVien00,BrayWinkler04}. Further studies have focused on the extreme value statistics of non-intersecting Brownian paths and bridges \cite{SchehrMaj08,RambeauSchehr10,Schehr12,Forrester11,Liechty12,Remenik17,Remenik172}.

In this work, we generalize the Ferrari-Spohn model \cite{FerrariSpohn05} to $N$ Brownian bridges conditioned not to cross each other, and to avoid a moving wall, see Fig. \ref{fig:IllustrationModel}. The main result of this paper is an expression for the joint distribution of their positions at the mid-time, in the form of a determinant of an $N\times N$ matrix whose entries are given in terms of the Airy function, see Eq.~\eqref{eq:ResDet} below. This joint distribution constitutes a determinantal point process, see Eq.~\eqref{eq:RNSingleDeterminant}, with a kernel given in Eq.~\eqref{eq:kernel_def}. Furthermore, we find that this distribution coincides with the joint distribution of the positions of $N$ fermions in the presence of a linear potential and a hard wall. Finally, we explore some of the features that emerge in the large-$N$ limit, such as the average spatial density of particles and their correlations, and the distributions of the positions of the top and bottom particles.

The structure of the paper is as follows. In section \ref{sec:mainResults} we give the precise definition of the problem that we solve, and give the main results and the main idea behind their derivation. The details of these derivations, including the mapping to the system of non-crossing Brownian particles under a square root barrier, as well as a more careful treatment of the boundary conditions in the definition of the problem,
are given in section \ref{sec:DetailedDerivation}. In section \ref{sec:determinantalStructure} we obtain the expression for the joint distribution of the positions, show the mapping to the fermion system, and then explore the large-$N$ behavior. We conclude and discuss our results in section \ref{sec:conclusion}.

\section{Presentation of the mapping and results}
 
 \label{sec:mainResults}

\subsection{Definition of the problem}

\begin{figure}[h!]
  \includegraphics[width=0.6\linewidth]{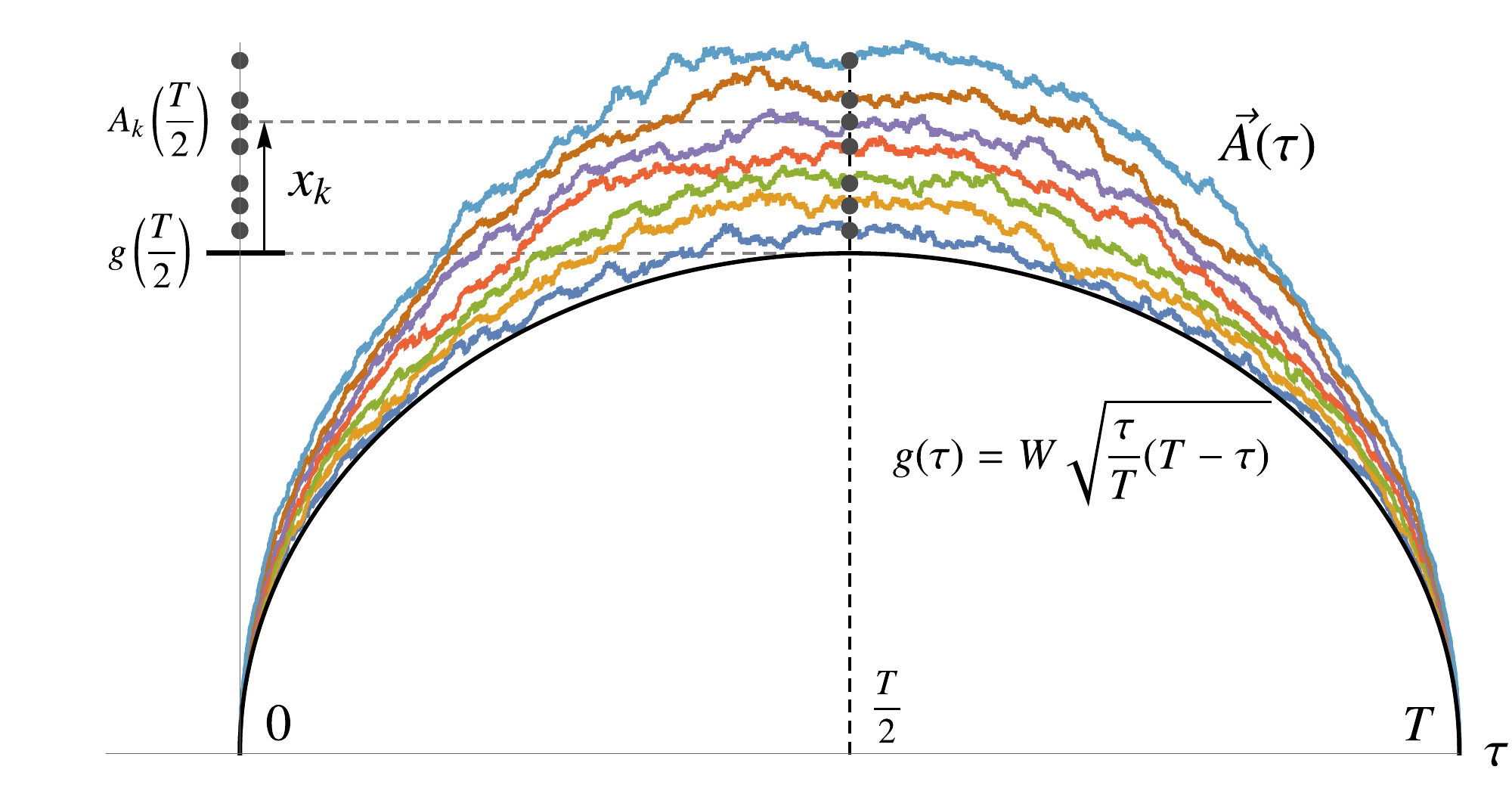}
  \caption{Illustration of a 7-particle realization of our model of $N$ non-intersecting Brownian bridges conditioned to stay above a moving wall. The dots indicate the positions of the particles at the observation time $\tau=T/2$.}
  \label{fig:IllustrationModel}
\end{figure}

Let $\vec{A} = (A_k )_{1 \leqslant k \leqslant N} $ be a collection of $N$ Brownian motions defined on the time interval $[0,T]$, started from $0$ at time $\tau=0$. We consider standard Brownian motions, i.e. such that $\mathbb{E} \left[ \left( A_k(\tau_1) - A_k(\tau_2) \right)^2 \right] = \tau_1 - \tau_2$  (for $\tau_1 \geqslant \tau_2$ and for all $k$), where $\mathbb{E}(\cdots \! \, )$ denotes the expectation value.
This amounts to setting the diffusion coefficient $D= \frac{1}{2}$. In the spirit of \cite{FerrariSpohn05}, we condition $\vec{A}$ on three events:
\begin{itemize}
\item First, $ A_k (T) = 0$ for all $k$, such that all paths end back at the origin at time $T$;
\item Second, $\forall \tau \in ]0,T[,  \ g(\tau)< A_k(\tau) $ for all $k$, such that all paths remain above the semicircle boundary $g$ on $]0,T[$, where $W>0$ is fixed and:
\begin{equation}
\label{eq:gDefinition}
g(\tau)= W \sqrt{\frac{\tau}{T}  (T- \tau ) } \; .
\end{equation}
\item  Third, $ \forall \tau \in ]0,T[, \  A_{k}(\tau) < A_{k+1}(\tau)$  for all $k \leqslant N-1$, such that all paths remain non-intersecting and ordered on $]0,T[$.
\end{itemize}
The particular case $N=1$ thus corresponds to the Ferrari-spohn model \cite{FerrariSpohn05}.
See Fig. \ref{fig:IllustrationModel} for an illustration of the model for $N=7$ particles.

We are interested in the probability density $R\left(\vec{x},W\right)$ for the paths $A_1, \ldots, A_N$ to be found at respective distances $x_1, \ldots, x_N$ from the barrier at mid-time $\frac{T}{2}$, such that for all $k$:
\begin{equation}
A_k \left( \frac{T}{2} \right) = g\left(\frac{T}{2} \right) + x_k  \; .
\end{equation}
For ease of notation, let us introduce some conventions. We denote by $\mathrm{B}^A_{[a,b]}$ the event that all paths remain over the barrier between times $a$ and $b$, that is $ \left\{ \forall\tau\in[a,b],\ \forall k\in\left\{ 1,\dots,N\right\} ,\ g(\tau)<A_{k}(\tau)\right\}  $. Similarly, we denote by $\mathrm{NC}^A_{[a,b]}$ the event that the paths remain non-crossing between times $a$ and $b$, $ \left\{ \forall\tau\in[a,b],\ \forall k\in\left\{ 1,\dots,N-1\right\} ,\ A_{k}(\tau)<A_{k+1}(\tau)\right\} $. Finally, we denote by $\mathrm{M}^A( \vec{x} )$ the event that all paths $A_k$ are respectively found at the distances $x_k$ from the barrier at mid-time, $\left\{ \forall k\in\left\{ 1,\dots,N\right\} ,\ A_{k}\left(\frac{T}{2}\right)=g\left(\frac{T}{2}\right)+x_{k}\right\} $. Let us sum up these notations:
\begin{eqnarray}
\mathrm{B}^A_{[a,b]} \quad &=& \quad \left\{ \forall\tau\in[a,b],\ \forall k\in\left\{ 1,\dots,N\right\} ,\ g(\tau)<A_{k}(\tau)\right\},  \\
\mathrm{NC}^A_{[a,b]} \quad &=& \quad \left\{ \forall\tau\in[a,b],\ \forall k\in\left\{ 1,\dots,N-1\right\} ,\ A_{k}(\tau)<A_{k+1}(\tau)\right\},  \\
\mathrm{M}^A( \vec{x} ) \quad &=& \quad  \left\{ \forall k\in\left\{ 1,\dots,N\right\} ,\ A_{k}\left(\frac{T}{2}\right)=g\left(\frac{T}{2}\right)+x_{k}\right\}. 
\end{eqnarray} 
With these notations, $R\left(\vec{x},W\right)$ can be written simply as the following conditional probability density, where we recall that the law of $\vec{A}$ is that of an $N$-dimensional Brownian motion process:
\begin{equation}
\label{eq:DefR}
R\left(\vec{x},W\right) = P\left(  \mathrm{M}^A( \vec{x} )  \mid \vec{A}(0) = \vec{A}(T) = \vec{0} \; , \; \mathrm{B}^A_{]0,T[}  \; , \; \mathrm{NC}^A_{]0,T[} \right) .
\end{equation}

In order to avoid a singularity in the initial and final conditions, we consider henceforth a situation where the processes start at a small non-zero initial time $0^+$ and end at time $T^-$ slightly smaller than $T$, and are found at these times in a position vector $\vec{0}^+$ that ensures the non-crossing conditions at the boundaries%
\footnote{We introduce the times $0^+$ and $T^-$ for technical reasons in order to enable us to use the results from \cite{NonCrossing}, as explained in detail in Section \ref{sec:DetailedDerivation}}.
The initial condition will also be denoted $\vec{A}(0^+) \approx \vec{0}$ when useful. As we show in full detail in Section \ref{sec:DetailedDerivation}, the conditional probability $R(\vec{x},W)$ has a well-defined limit. A similar regularization of initial and final conditions is usually performed whenever considering Brownian excursions, as well as in the original Ferrari-Spohn model.
 
\subsection{Mapping presentations}
 
With the notations introduced in the last paragraph, we are looking to calculate the conditional probability:
\begin{equation}
\label{eq:DefRNewNotations}
R\left(\vec{x},W\right) = P\left(  \mathrm{M}^A( \vec{x} )  \mid \vec{A}(0^+) = \vec{A}(T^-) = \vec{0}^+ \; , \; \mathrm{B}^A_{[0^+,T^-]}  \; , \; \mathrm{NC}^A_{[0^+,T^-]} \right) .
\end{equation}
The crux of our solution is to transform this problem into one concerning a collection of non-crossing Brownian motions over a barrier with a square root profile, such that results from \cite{NonCrossing} apply directly. In order to do this, a classical mapping between Brownian motions with a square root barrier and Brownian bridges with a semicircle barrier is employed. In this section we present a bird's eye view of the derivation, with the details given in the next section.

Before we introduce this mapping, let us slightly reshape the problem. It is showed in full detail in the next section that $R\left(\vec{x},W\right)$ can be written as follows:

\begin{equation}
\label{eq:ReshapedProblemDef}
R\left(\vec{x},W\right) = \frac{
P\left(  \mathrm{M}^A( \vec{x} )  ,   \mathrm{B}^A_{[0^+,\frac{T}{2}]}  ,  \mathrm{NC}^A_{[0^+,\frac{T}{2}]}  \mid  \vec{A}(0^+) \approx \vec{A}(T) = \vec{0} \right) ^2
}{P\left(   \vec{A}(T^-) = \vec{0}^+ ,  \mathrm{B}^A_{[0^+,T^-]}  ,  \mathrm{NC}^A_{[0^+,T^-]}  \mid \vec{A}(0^+) = \vec{0}^+ \right)}
\frac{
P \left(\vec{A}(T) = \vec{0} \mid \vec{A}(0^+) = \vec{0}^+ \right)^2
}{
P \left( \vec{A}(T) = \vec{0} \mid M^A(\vec{x}) \right)^2
} \; .
\end{equation}
The term which appears in the numerator of the left fraction,
\be
\label{ProbTerm}
Q(\vec{x},W)=
P\left(  \mathrm{M}^A( \vec{x} )  ,   \mathrm{B}^A_{[0^+,\frac{T}{2}]}  ,  \mathrm{NC}^A_{[0^+,\frac{T}{2}]}  \mid  \vec{A}(0^+) \approx \vec{A}(T) = \vec{0} \right) ,
\ee
is a probability involving $N$ Brownian bridges: the $N$ particles are started from $0$ and are conditioned to end back at $0$ at $\tau=T$. More precisely, this term is the probability for the $N$ Brownian bridges to be found at positions given by $\mathrm{M}( \vec{x} )$ at $\tau = \frac{T}{2}$ and to stay above the semicircle barrier while remaining non-crossing on $[0^+, \frac{T}{2}]$.

We now introduce the mapping that transforms this probability into one concerning Brownian motions over a square root barrier. This mapping was used in section V of \cite{NonCrossing} in a different setting. The following change of variables maps a Brownian bridge $A(\tau)$ defined for $\tau \in [0,T]$ and conditioned to end back at $0$ at $\tau = T$, to a standard Brownian motion $B(t)$ with a time variable $t =\frac{T \tau}{T-\tau} \in [0,\infty[$, see \cite{RevuzYor}:
\begin{equation}
\label{eq:AToBMapping}
A(\tau) 
\quad  \xrightarrow{ \ \;  \tau \ \;  \to \ \; t =\frac{T \tau}{T-\tau} \ \;  } \quad
B(t) =   \frac{T+t}{T} \; A\left(\frac{T t }{T+t}\right) 
\end{equation}
The semicircle barrier $g(\tau)$ is mapped under this transformation to the square root barrier $h(t)$:
\begin{equation}
\label{eq:gTohMapping}
g(\tau) =   W \sqrt{\frac{\tau}{T}  (T- \tau ) }
\quad  \xrightarrow{ \quad \quad \quad \quad} \quad
h(t) = W \sqrt{t} .
\end{equation}
The inverse mapping is given by:
\begin{equation}
B(t) 
\quad  \xrightarrow{ \ \;  t \ \;  \to \ \; \tau =\frac{T t}{T+t} \ \;  } \quad
A(\tau) = \frac{T- \tau}{T} B\left(\frac{T\tau}{T-\tau}\right).
\end{equation}
We are particularly interested in the mid-time $\tau = \frac{T}{2}$, where the following relations hold:
\begin{equation}
\tau = \frac{T}{2} \quad \quad  \implies \quad \quad 
\left\{
    \begin{array}{ll}
        t = T , \\[3pt]
        A(\frac{T}{2}) = \frac{1}{2} \, B(T) , \\[3pt]
        g(\frac{T}{2}) = \frac{1}{2} \,   h(T) =   W\frac{\sqrt{T}}{2} .
    \end{array}
\right.
\label{eq:MidTimeRelations}
\end{equation}
The mapping is illustrated in Fig. \ref{fig:mapping} for $N=7$ particles.

Applying this mapping to each of the $N$ paths $A_k(\tau) \longrightarrow B_k(t)$ and to the barrier $ g(\tau) \to h(t)$, we note that the mapping conserves the non-crossing property of the $N$-particle path, between particles and with the barrier. We thus have the following equivalences:
\begin{eqnarray}
\mathrm{B}^A_{[0^+,\frac{T}{2}]} 
&\iff & \mathrm{B}^B_{[0^+,T]}  =
 \big\{  \forall t \in [0^+,T] , \ \forall  k \leqslant N, \   h(t) < B_k(t)   \big\}  \, ,
\\
\mathrm{NC}^A_{[0^+,\frac{T}{2}]} 
&\iff &
\mathrm{NC}^B_{[0^+,T]}  =
 \big\{   \forall t \in [0^+,T], \ \forall  k \leqslant N-1, \  B_{k}(t) < B_{k+1}(t)       \big\} \, .
\end{eqnarray}
Here we have introduced events $\mathrm{B}^B_{[a,b]} $ and $\mathrm{NC}^B_{[a,b]}$ for the set of Brownian motion $B$ processes, defined in an analogous fashion as the previously defined $\mathrm{B}^A_{[a,b]} $ and $\mathrm{NC}^A_{[a,b]}$ involving Brownian bridge $A$ processes. Similarly, we denote by $M^B(\vec{x})$ the event that at $t=T$, corresponding to $\tau = \frac{T}{2}$, the $B_k$ processes are found at respective distances $x_k$ from the square root barrier $h$. From the mid-time relations \eqref{eq:MidTimeRelations}, we have:
\begin{equation}
\mathrm{M}^A(\vec{x})
\iff\mathrm{M}^{B}(2\vec{x})=\left\{ \forall k\in\left\{ 1,\dots,N\right\} ,\ B_{k}\left(T\right)=h\left(T\right)+2x_{k}\right\}.
\end{equation}
We can thus directly write the probability of interest \eqref{ProbTerm} involving Brownian bridges in terms of a probability involving Brownian motions, with a Jacobian factor $2^N$ from \eqref{eq:MidTimeRelations}:
\begin{equation}
\label{eq:ProbTermMappedToB}
Q(\vec{x},W)=
P\left(  \mathrm{M}^A( \vec{x} )  ,   \mathrm{B}^A_{[0^+,\frac{T}{2}]}  ,  \mathrm{NC}^A_{[0^+,\frac{T}{2}]}  \mid  \vec{A}(0^+) \approx \vec{A}(T) = \vec{0} \right) =
 2^N  P\left(  \mathrm{M}^B(2 \vec{x} )  \ , \  \mathrm{B}^B_{[0^+,T]}  \ , \  \mathrm{NC}^B_{[0^+,T]}  \mid  \vec{B}(0^+) =  \vec{0}^+ \right).
\end{equation}

The interest of this effort is that the problem has been transformed to the computation of the probability that $N$ independent Brownian motions do not cross each other and stay above the square root barrier $h(t)$ until time $t=T$, and that they are found at a set of given positions at time $T$. 
This is exactly what was computed in \cite{NonCrossing}, by using the quantum generator of a system of $N$ noninteracting spinless fermions in a quadratic potential with a hard-wall located at $W$, the barrier parameter. Denoting by $X_1,\dots,X_N$ the fermions' positions, the total Hamiltonian $\hat{\mathcal{H}}$ of this system of independent fermions is the sum of the 1-particle Hamiltonians $\hat{H}^{(k)}$, obtained through $X \to X_k$ from the generic 1-particle Hamiltonian $\hat{H}$:
\begin{equation}
\label{Hamiltonian}
\hat{\mathcal{H}}
=\sum_{k=1}^{N}\hat{H}^{(k)},
\qquad
\hat{H}=\begin{cases}
-\frac{1}{2}\frac{\partial^{2}}{\partial X^{2}}+\frac{1}{8}X^{2}-\frac{1}{4} \quad  & X>W,\\[1mm]
+\infty  & X\leqslant W.
\end{cases}
\end{equation}
We denote by $(\psi_i)_{i\geqslant 0}$ the sequence of 1-particle eigenfunctions of $\hat{H}$, and $(E_i)_{i \geqslant 0}$ the corresponding sequence of eigenvalues. The dependence of both sequences on $W$ is not made explicit for ease of notation. We give the expression of $\psi_i$, constrained to vanish at $X = W$ and $X \to \infty$, in terms of the parabolic cylinder function $D_p(z)$, see \cite{NonCrossing}:
\begin{equation}
\label{eq:ExpressionPsiParabolicCylinder}
\psi_i ( X ) = C_i  \; D_{2 E_i} (X)\mathbbm{1}_{W<X} \; .
\end{equation}
Here $\mathbbm{1}_{(\cdots)}$ is the indicator function which takes the value $1$ if the condition $\cdots$ is satisfied, and zero otherwise,
 the corresponding eigenvalue $E_i$ is constrained to be the $i+1$-th root in increasing order of the eigenvalue equation $D_{2E_i}(W) =0$, and $C_i$ is a normalization constant. The $N$-particle eigenfunctions of $\hat{\mathcal{H}}$ are obtained as Slater determinants of $\psi$ functions and are indexed by $\vec{k} \in \Omega_{N}$ where $\Omega_{N}=\left\{\vec{k} \in \mathbb{N}^{N} \text { such that } k_{1}<k_{2}<\cdots<k_{N}\right\}$. The corresponding $N$-particle eigenvalues are $E_{\vec{k}}(N,W)= \sum\limits_{i=1}^N E_{k_i}$.

The results from \cite{NonCrossing} allow to express $Q(\vec{x},W)$ exactly in terms of $\psi$ functions and corresponding eigenvalues, in a way that depends on initial conditions and is left to the next section, see Eq.~\eqref{eq:QuantumResultQ}. We will focus on the large $W$ limit where $\psi_k$ becomes:
\begin{equation}
\label{eq:ResultPsiLargeW}
\psi_{k}(X) \simeq D_{k,W}\,\operatorname{Ai}\left(\alpha_{k+1} + \left(\frac{W}{2}\right)^{1/3}(X-W)\right)  \mathbbm{1}_{W<X}
\end{equation}
where $(\alpha_k)_{k \geqslant 1}$ is the sequence of zeros of the Airy function $\text{Ai}(x)$, e.g. $\alpha_1=-2.33811\ldots$, and $D_{k,W}$ is a constant that is calculated below, and ensures the normalization $\int_{W}^{\infty}\left[\psi_{k}\left(X\right)\right]^{2}dX=1$.

The three other terms in \eqref{eq:ReshapedProblemDef} are easily dealt with: $ P\left(\vec{A}(T) = \vec{0} \mid \vec{A}(0^+) = \vec{0}^+ \right)$ and $ P \left( \vec{A}(T) = \vec{0} \mid M^A(\vec{x}) \right)$ are Brownian transition probabilities and $P\left(   \vec{A}(T^-) = \vec{0}^+ ,  \mathrm{B}^A_{[0^+,T^-]}  ,  \mathrm{NC}^A_{[0^+,T^-]}  \mid \vec{A}(0^+) = \vec{0}^+ \right)$ is a constant with respect to the $x$ variables and therefore contributes only to the overall normalization factor.

\subsection{Results}

The computation described in the previous subsection yields the desired distribution $R\left(\vec{x},W\right) $, in the large $W$ limit, as the squared determinant of a matrix whose entries are given in terms of the Airy function:
\begin{equation} 
\label{eq:ResDet}
R\left(\vec{x},W\right) = \frac{ (4W)^{\frac{N}{3}}
}{
T^{\frac{N}{2}} \
\prod_{k=1}^N \operatorname{Ai}^\prime (\alpha_k)^2
} \; 
 \left[ \; \det_{1\leqslant i,j \leqslant N}
\operatorname{Ai} \left(  \alpha_{i}  + \frac{\left(4 W \right)^{\frac{1}{3}}}{\sqrt{T}}  \ x_j  \right) 
\right]^2  \mathbbm{1}_{0<x_1<\ldots<x_N}\;.
\end{equation}
In the particular case $N=1$ this becomes simply the Ferrari-Spohn distribution \cite{FerrariSpohn05, GeomOpt3}
\begin{equation}
\rho(x)=\frac{\left(4W\right)^{1/3}}{\sqrt{T}\operatorname{Ai}'(\alpha_{1})^{2}}\operatorname{Ai}^{2}\left(\alpha_{1}+\frac{\left(4W\right)^{1/3}}{\sqrt{T}}\ x\right) \mathbbm{1}_{0<x}\;.
\end{equation}
The coefficient $\ell=\left(4W\right)^{1/3} \! /\sqrt{T}$ in front of $x$ in the Airy function agrees with equation (15) of \cite{GeomOpt3}. Indeed, this distribution is encountered under fairly general assumptions, where the coefficient is in general given by $\ell=\left(\frac{-g''\left(\tau_{\text{obs}}\right)}{2D^{2}}\right)^{1/3}$ (and with a corresponding normalization) \cite{GeomOpt3}. The particular case considered here corresponds to the semicircle boundary \eqref{eq:gDefinition}, the diffusion coefficient $D=1/2$, and the observation time $\tau_{\text{obs}}=T/2$.

The distribution \eqref{eq:ResDet} can be written as a single determinant
\begin{equation}
\label{eq:RNSingleDeterminant}
R\left(\vec{x},W\right)=N!\det_{1\leq i,j\leq N}K_{N}\left(x_{i},x_{j}\right)
\end{equation}
with the kernel
\begin{equation}
\label{eq:kernel_def}
K_{N}\left(x,x'\right)=\sum_{k=1}^{N}\phi_{k}\left(x\right)\phi_{k}\left(x'\right)
\end{equation}
and where the functions $\phi_k$ are defined, for $k \geqslant 1$, as 
\be
\label{eq:phidef} 
\phi_{k}\left(x\right)=\frac{\left(4W\right)^{1/6}}{T^{1/4}\left|\text{Ai}'\left(\alpha_{k}\right)\right|}\text{Ai}\left(\alpha_{k}+\frac{(4W)^{1/3}}{\sqrt{T}}x\right)  \mathbbm{1}_{0<x} \, ,\quad k=1,2,\dots\;.
\ee

These expressions of $R(\vec{x},W)$ are obtained below with all details, in particular regarding the boundary conditions and their limit. The determinantal structure of the distribution is exploited in section \ref{sec:determinantalStructure} to explore the large-N behavior.

\section{Detailed derivation}
\label{sec:DetailedDerivation}

In this section, we detail the procedures and computations that were briefly brushed in the previous section. We begin by regularizing the initial and final conditions, in order to avoid the problem that the conditioning event $\{\vec{A}(0) = \vec{A}(T) = \vec{0} \; , \; \mathrm{B}^A_{]0,T[}  \; , \; \mathrm{NC}^A_{]0,T[}\}$ has zero probability.
The initial and final positions of the $N$ Brownian motions $\vec{A}$ will now be taken to be a vector $\vec{s}$, such that $0 < s_1 < \cdots < s_N$. Furthermore, the results from \cite{NonCrossing} impose a starting time that is not exactly zero, therefore we introduce a time $\tau_0$ at which the Brownian motions are started, and the final position is now observed at $T-\tau_0$ by symmetry. 
The initial and final conditions are thus given by:
\begin{equation}
\vec{A}(\tau_0)=\vec{A}(T-\tau_0)= \vec{s} .
\end{equation}
With these precise conditions, the problem that we now solve is to evaluate the probability density $R\left(\vec{x},W\right)$ for a set of $N$ standard Brownian motions $\vec{A}$:
\begin{equation}
R\left(\vec{x},W\right) = P\left(  \mathrm{M}^A( \vec{x} )  \mid \vec{A}(\tau_0) = \vec{A}(T-\tau_0) = \vec{s} \; , \; \mathrm{B}^A_{[\tau_0,T-\tau_0]}  \; , \; \mathrm{NC}^A_{[\tau_0,T-\tau_0]} \right).
\end{equation}
We stress that, for ease of notation, we do not write explicitly the dependence of $R\left(\vec{x},W\right)$ on the $\tau_0$ and $\vec{s}$ variables. The limits $\tau_0 \to 0$ and $\vec{s} \to \vec{0}$ are handled at the end of the section.

\subsection{Mapping to Brownian bridges over the semicircle barrier on $[\tau_0,\frac{T}{2}]$}

Let us rewrite the distribution $R(\vec{x},W)$ in terms of a probability density involving Brownian bridges, so as to obtain equation \eqref{eq:ReshapedProblemDef} in the previous section. The conditional probability $R\left(\vec{x},W\right)$ can be written as:
\begin{equation}
R\left(\vec{x},W\right) = \frac{P\left(  \mathrm{M}^A( \vec{x} )  , \vec{A}(T-\tau_0) = \vec{s}  , \mathrm{B}^A_{[\tau_0,T-\tau_0]}  ,  \mathrm{NC}^A_{[\tau_0,T-\tau_0]}  \mid  \vec{A}(\tau_0) = \vec{s} \right)  }{P\left(   \vec{A}(T-\tau_0) = \vec{s} ,  \mathrm{B}^A_{[\tau_0,T-\tau_0]}  ,  \mathrm{NC}^A_{[\tau_0,T-\tau_0]}  \mid \vec{A}(\tau_0) = \vec{s} \right)}
\end{equation}
By Markov's property, the numerator can be cut in two terms corresponding to $[\tau_0, \frac{T}{2}]$ and $[\frac{T}{2},T-\tau_0]$. Moreover, these terms are equal by symmetry:
\begin{eqnarray}
R\left(\vec{x},W\right) &=& \frac{
P\left(  \mathrm{M}^A( \vec{x} )   ,   \mathrm{B}^A_{[\tau_0,\frac{T}{2}] }  ,  \mathrm{NC}^A_{[\tau_0,\frac{T}{2}]}  \mid  \vec{A}(\tau_0) = \vec{s} \right)  
P\left(   \vec{A}(T-\tau_0) = \vec{s}  , \mathrm{B}^A_{[\frac{T}{2},T-\tau_0]}  ,  \mathrm{NC}^A_{[\frac{T}{2},T-\tau_0]}  \mid  \mathrm{M}^A( \vec{x} )  \right)  
}{P\left(   \vec{A}(T-\tau_0) = \vec{s} ,  \mathrm{B}^A_{[\tau_0,T-\tau_0]}  ,  \mathrm{NC}^A_{[\tau_0,T-\tau_0]}  \mid \vec{A}(\tau_0) = \vec{s} \right)} \nn\\
 &=& \frac{
P\left( \mathrm{M}^A( \vec{x} )   ,   \mathrm{B}^A_{[\tau_0,\frac{T}{2}]}  ,  \mathrm{NC}^A_{[\tau_0,\frac{T}{2}]}  \mid  \vec{A}(\tau_0) = \vec{s} \right) ^2
}{P\left(   \vec{A}(T-\tau_0) = \vec{s} ,  \mathrm{B}^A_{[\tau_0,T-\tau_0]}  ,  \mathrm{NC}^A_{[\tau_0,T-\tau_0]}  \mid \vec{A}(\tau_0) = \vec{s} \right)} \; .
\label{eq:Intermediate}
\end{eqnarray}
In order to obtain a probability concerning a set of Brownian bridges, we now need to introduce back a conditioning on the final position, and more specifically to impose that $\vec{A}(T)=\vec{0}$. This is not problematic at this point of the computation, because the non-crossing events will now only concern the $\tau \in [\tau_0,\frac{T}{2}]$ time window, such that the particles can be harmlessly supposed to all end at $0$ at $\tau = T$. We then write the following relations on the numerator term:
\begin{eqnarray}
 P  \bigg( \mathrm{M}^A( \vec{x} )   , &  \mathrm{B}^A_{[\tau_0,\frac{T}{2}]}  &  ,  \mathrm{NC}^A_{[\tau_0,\frac{T}{2}]}  \mid  \vec{A}(\tau_0) = \vec{s} \bigg)     P \left( \vec{A}(T) = \vec{0} \mid M^A(\vec{x}) \right) \nn\\
&=& 
P\left( \mathrm{M}^A( \vec{x} )   ,   \mathrm{B}^A_{[\tau_0,\frac{T}{2}]}  ,  \mathrm{NC}^A_{[\tau_0,\frac{T}{2}]}  , \vec{A}(T) = \vec{0} \mid  \vec{A}(\tau_0) = \vec{s} \right) 
\nn\\
&=& 
P\left(  \mathrm{M}^A( \vec{x} )  ,   \mathrm{B}^A_{[\tau_0,\frac{T}{2}]}  ,  \mathrm{NC}^A_{[\tau_0,\frac{T}{2}]}  \mid  \vec{A}(\tau_0) = \vec{s} , \vec{A}(T) = \vec{0} \right)   
P \left(\vec{A}(T) = \vec{0} \mid \vec{A}(\tau_0) = \vec{s} \right) \, .
\end{eqnarray}
Injecting this relation in \eqref{eq:Intermediate} yields the desired Eq.~\eqref{eq:ReshapedProblemDef}, rewritten here with our regularized initial and final conditions:
\begin{equation} 
R\left(\vec{x},W\right) = \frac{
P\left(  \mathrm{M}^A( \vec{x} )  ,   \mathrm{B}^A_{[\tau_0,\frac{T}{2}]}  ,  \mathrm{NC}^A_{[\tau_0,\frac{T}{2}]}  \mid  \vec{A}(\tau_0) = \vec{s} , \vec{A}(T) = \vec{0}  \right) ^2
}{P\left(   \vec{A}(T-\tau_0) = \vec{s} ,  \mathrm{B}^A_{[\tau_0,T-\tau_0]}  ,  \mathrm{NC}^A_{[\tau_0,T-\tau_0]}  \mid \vec{A}(\tau_0) = \vec{s} \right)}
\frac{
P \left(\vec{A}(T) = \vec{0} \mid \vec{A}(\tau_0) = \vec{s} \right)^2
}{
P \left( \vec{A}(T) = \vec{0} \mid M^A(\vec{x}) \right)^2
} \, .
\label{eq:ReshapedProblemDefDetailedDerivation}
\end{equation} 
The term in the denominator of the fraction is simply given by the propagator of the diffusion equation:
\begin{equation}
\label{eq:AGivenMA}
 P \left( \vec{A}(T) = \vec{0} \mid M^A(\vec{x}) \right) = \frac{1}{( \pi T )^{\frac{N}{2}}}  \ e^{-\frac{1}{T} \sum\limits_{k=1}^N  \left( g(\frac{T}{2}) +  x_k - s_k \right) ^2   } \, .
 \end{equation} 
  
As explained in section \ref{sec:mainResults}, the interest of this calculation is to introduce a probability concerning $N$ Brownian bridges, i.e., Brownian motions conditioned on the event $\vec{A}(T)=\vec{0}$. More specifically, the term we obtain is the probability density that the particles are found at positions $g(\frac{T}{2}) +  \vec{x}$ at the middle-time $\frac{T}{2}$, and that they stay non-crossing and above the semicircle barrier on $[\tau_0,\frac{T}{2}]$. As in \eqref{ProbTerm}, we denote it $Q(\vec{x},W)$:
\begin{equation}
\label{eq:QdefRegularized}
Q(\vec{x},W) = P\left(  \mathrm{M}^A( \vec{x} )  ,   \mathrm{B}^A_{[\tau_0,\frac{T}{2}]}  ,  \mathrm{NC}^A_{[\tau_0,\frac{T}{2}]}  \mid  \vec{A}(\tau_0) = \vec{s} , \vec{A}(T) = \vec{0}  \right) \, .
\end{equation}
In the same fashion as $R(\vec{x},W)$, we suppress the dependence of $Q(\vec{x},W)$ on the initial condition. We note that $R (\vec{x},W)$ is then, by injecting here $g(T/2) = W\sqrt{T}/2$, and by using Eqs.~\eqref{eq:AGivenMA} and \eqref{eq:QdefRegularized} in \eqref{eq:ReshapedProblemDefDetailedDerivation},
\begin{equation}
\label{eq:RasFunctionOfQ}
R (\vec{x},W) = Q(\vec{x},W)^2 \  e^{\frac{2}{T} \sum\limits_{k=1}^N  \left( \frac{W\sqrt{T}}{2} +  x_k - s_k \right) ^2   } \frac{ (\pi T)^N \ P \left(\vec{A}(T) = \vec{0} \mid \vec{A}(\tau_0) = \vec{s} \right)^2}{P\left(   \vec{A}(T-\tau_0) = \vec{s} ,  \mathrm{B}^A_{[\tau_0,T-\tau_0]}  ,  \mathrm{NC}^A_{[\tau_0,T-\tau_0]}  \mid \vec{A}(\tau_0) = \vec{s} \right)} \, .
\end{equation}

\subsection{Mapping to Brownian motions over a square root barrier}

The key step that we now take is to transform the problem concerning Brownian bridges over a semicircle boundary to a problem concerning Brownian motions over a square root boundary. 
For the convenience of the reader, we write again the change of variables, given in Eqs.~\eqref{eq:AToBMapping} and \eqref{eq:gTohMapping}, that maps a Brownian bridge $A(\tau)$ defined for $\tau \in [0,T]$ (and conditioned to end back at $0$ at $\tau = T$), to a standard Brownian motion $B(t)$ with a time variable $t =\frac{T \tau}{T-\tau} \in [0,\infty[$ \cite{RevuzYor}:
\begin{equation} 
A(\tau)\quad\to\quad B\left(t\right)=\frac{T+t}{T}\;A\left(\frac{Tt}{T+t}\right),\qquad\qquad g(\tau)=W\sqrt{\frac{\tau}{T}(T-\tau)}\quad\to\quad h\left(t\right)=W\sqrt{t}.
\end{equation}  
We are particularly interested in the mid-time $\tau = T/2$, where the relations \eqref{eq:MidTimeRelations} hold. Under this mapping, the initial condition $\vec{A}(\tau_0) = \vec{s}$ is mapped to
\begin{equation}
\label{eq:InitialConditionBtzero} 
\vec{B}(t_{0})=\frac{T}{T-\tau_{0}}\vec{s}\quad \text{where} \quad t_{0}=\frac{T\tau_{0}}{T-\tau_{0}}.
\end{equation}   

From the equivalences $ {\mathrm{B}^A_{[\tau_0,\frac{T}{2}]}   \iff  \mathrm{B}^B_{[t_0,T]}}  $,  ${\mathrm{NC}^A_{[\tau_0,\frac{T}{2}]}  \iff  \mathrm{NC}^B_{[t_0,T]} } $ and ${\mathrm{M}^A(\vec{x})  \iff \mathrm{M}^{B}(2\vec{x}) }$ detailed in the previous section, we can then write Eq.~\eqref{eq:ProbTermMappedToB} given here with the regularized initial condition:
\begin{equation} 
Q(\vec{x},W)  =
 2^N  P\left(  \mathrm{M}^B(2 \vec{x} )  \ , \  \mathrm{B}^B_{[t_0,T]}  \ , \  \mathrm{NC}^B_{[t_0,T]}  \mid  \vec{B}(t_0) = \frac{T  }{T- \tau_0} \vec{s} \right).
\end{equation}

\begin{figure}[h!]
  \includegraphics[width=0.4\linewidth]{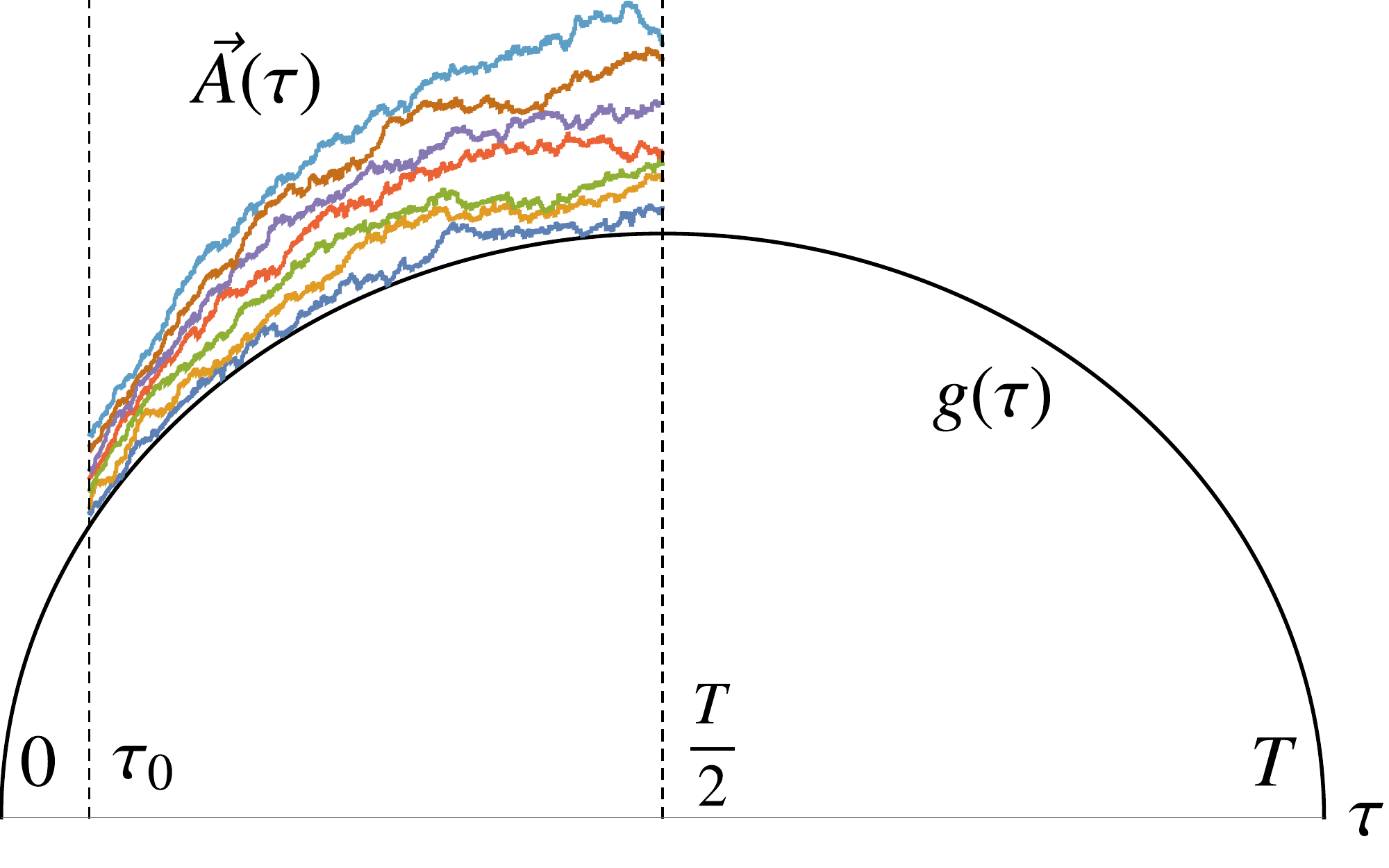}
  \hspace{0.02\linewidth}
  \includegraphics[width=0.4\linewidth]{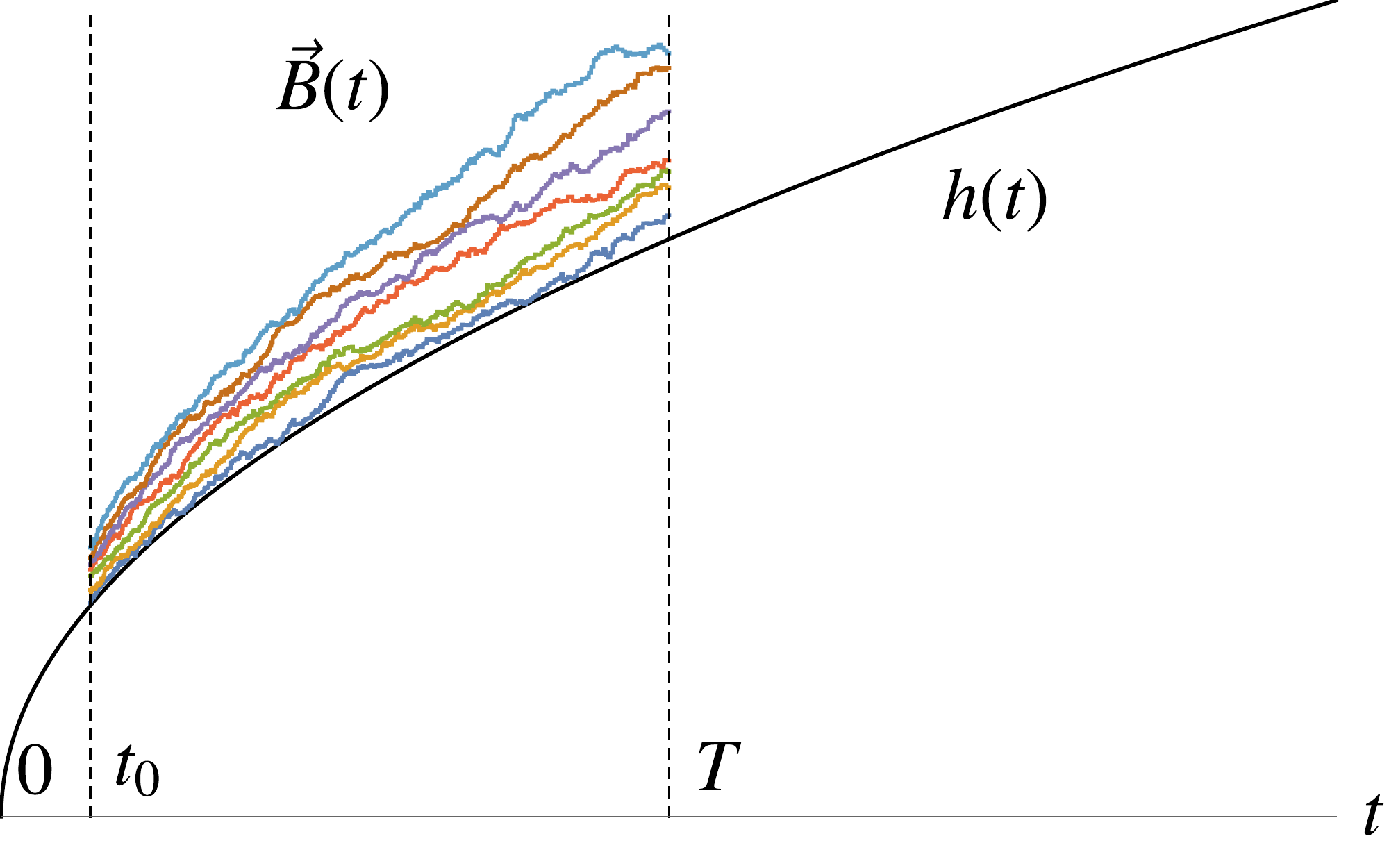}
  \caption{
 Illustration of the mapping from (left) 7 non-crossing Brownian bridges conditioned to stay over a semicircle barrier on $[\tau_0,\frac{T}{2}]$ to (right) 7 non-crossing Brownian motions conditioned to stay over a square root barrier on $[t_0, T]$, see Eqs.~\eqref{eq:AToBMapping} and \eqref{eq:gTohMapping}. Note that the right figure is scaled down by a factor 2 compared to the left figure.
}
  \label{fig:mapping}
\end{figure}

The problem has been transformed to the computation of the probability density for Brownian motions to be found at positions $h(T) + 2 \vec{x}$ at time $t=T$ and to have stayed non-crossing and above the boundary $h(t)=W \sqrt{t}$ on the time frame $t \in [t_0, T]$, with a given initial condition $\vec{B}(t_0)$, see an illustration of the mapping in Fig. \ref{fig:mapping}. This quantity was computed in \cite{NonCrossing}.
 With $(\psi_i)_{i\geqslant 0}$ the sequence of 1-particle eigenfunctions of the 1-particle hamiltonian $\hat{H}$ defined in \eqref{Hamiltonian} and $(E_i)_{i\geqslant 0}$ the corresponding eigenvalues, equation (48) in \cite{NonCrossing} gives%
\footnote{Note that Ref.~\cite{NonCrossing} uses different notations to this work and in particular that, in equation (48) of \cite{NonCrossing}, $T$ is not the constant of this work but the transformed time variable $\ln(t)$. Injecting in this equation the relations \eqref{eq:MidTimeRelations} and \eqref{eq:InitialConditionBtzero} yields \eqref{eq:QuantumResultQ}.}:
\begin{eqnarray}
\label{eq:QuantumResultQ}
Q(\vec{x},W)  &=&  \frac{2^{N}}{ T^{N/2}   } \,
e^{-\frac{1}{4}\sum\limits_{k=1}^N\left[  \left(W+\frac{2x_{k}}{\sqrt{T}}\right)^{2} -  \frac{T\tau_{0}}{T-\tau_{0}}{s}_{k}^{2}\right]} 
\nn\\
& \times &\sum_{\vec{k}\in\Omega_{N}}
\left[ 
\det\limits_{1 \leqslant i,j \leqslant N}\psi_{k_{i}}\left(W+2\frac{x_{j}}{\sqrt{T}}\right) \
\det\limits_{1 \leqslant i,j \leqslant N}\psi_{k_{i}}^{*}\left(\sqrt{\frac{T  }{(T-\tau_{0}) \tau_0 }} \, {s}_{j}\right) \ \left(\frac{T-\tau_{0}}{\tau_{0}}  \right) ^{- E_{\vec{k}}(N,W)}
\right] \, ,
\end{eqnarray}
where $\Omega_{N}=\left\{\vec{k} \in \mathbb{N}^{N} \text { such that } k_{1}<k_{2}<\cdots<k_{N}\right\}$ and $E_{\vec{k}}(N,W)= \sum\limits_{i=1}^N E_{k_i}$.

\subsection{Initial condition limit $\{ \tau_0, \vec{s} \} \to \{ 0 , \vec{0}\}$}

\begin{figure}[h!] 
  \includegraphics[width=0.4\linewidth]{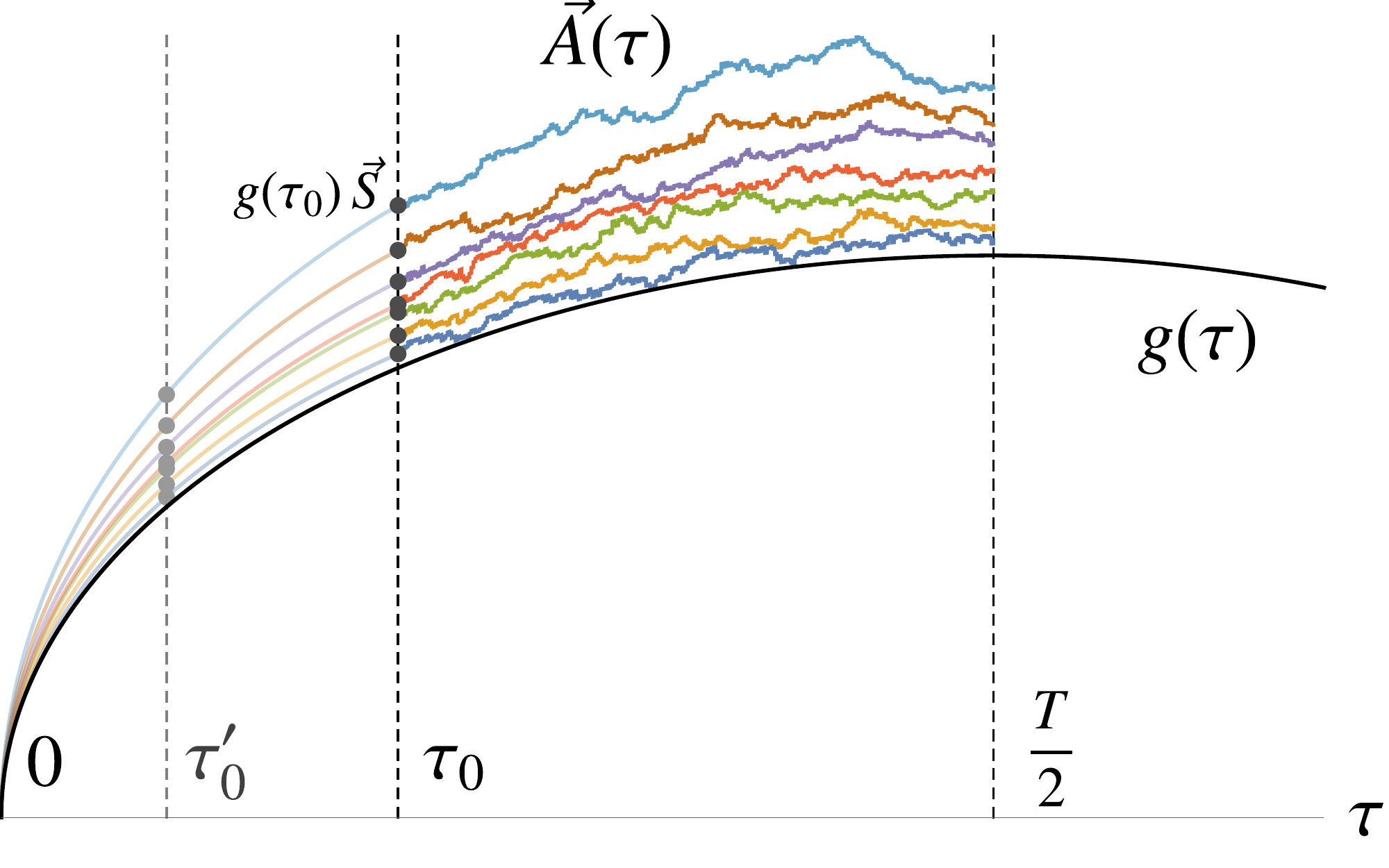}
  \caption{ 
Illustration of the regularization limit $\{ \tau_0, \vec{s} \} \to \{ 0 , \vec{0}\}$ as $\vec{s}(\tau_0) = g(\tau_0) \vec{S}$ sketched for a given $\vec{S}$ at two times $\tau_0' < \tau_0$.}
\label{fig:InitialLimit}
\end{figure}

In order to recover the results announced in section \ref{sec:mainResults} for Brownian bridges started and ended at exactly $0$, we must take a limit over the initial condition and to push $\tau_0$ to $0$ and $\vec{s}$ to $\vec{0}$. In order for this procedure to make sense, the non-crossing conditions must be respected by the particles $\vec{A}$ at starting positions $\vec{s}$ at time $\tau_0$:
\begin{equation}
g(\tau_0) = W \sqrt{\frac{\tau_0 }{T}  (T- \tau_0 ) }  \ <  \ s_1 \ < \  \cdots  \ < \ s_N .
\end{equation}
We do this by scaling $\vec{s}$ with $\tau_0$ as follows:
\be
\label{eq:Sdef}
\vec{s}\left(\tau_{0}\right)=g\left(\tau_{0}\right)\vec{S}=W \sqrt{\frac{\tau_{0}}{T}(T-\tau_{0})}\,\vec{S},
\ee
where $\vec{S}$ satisfies $1<S_{1}<\cdots<S_{N}$. We now take the limit $\tau_0 \to 0$ with $\vec{S}$ held constant. This limit procedure is sketched in Fig. \ref{fig:InitialLimit}.

Let us rewrite $Q(\vec{x},W)$, by injecting this initial condition scaling  \eqref{eq:Sdef} into \eqref{eq:QuantumResultQ}:
\be
Q(\vec{x},W)  =  \frac{2^{N}}{ T^{N/2}   }
e^{-\frac{1}{4}\sum\limits_{k=1}^N\left[  \left(W+\frac{2x_{k}}{\sqrt{T}}\right)^{2} -  \tau_0^2 W^2 S_{k}^{2} \right]} 
\!\! \sum_{\vec{k}\in\Omega_{N}}
\! \left[ 
\det\limits_{1 \leqslant i,j \leqslant N}\psi_{k_{i}} \! \left(W+2\frac{x_{j}}{\sqrt{T}}\right)
\! \det\limits_{1 \leqslant i,j \leqslant N}\psi_{k_{i}}^{*} \! \left(W S_{j}  \right) \! \left(\frac{T-\tau_{0}}{\tau_{0}}  \right) ^{- E_{\vec{k}}(N,W)}
\right]  .
\ee
Since $E_{\vec{k}}(N,W) \geqslant 0$, we see that the dominant contribution to the sum when taking the limit $\tau_0 \to 0$ is the ground state $\vec{k}_0=(0,1,\cdots,N-1)$. The initial condition term in the exponential furthermore vanishes and $T-\tau_0 \simeq T$, such that $Q(\vec{x},W)$ is equivalent in the $\tau_0 \to 0$ limit to:
\begin{eqnarray}
Q(\vec{x},W) 
\simeq \frac{2^{N}}{ T^{N/2}   } \,
e^{-\frac{1}{T}\sum\limits_{k=1}^N   \left( \frac{W \sqrt{T}}{2}+ x_{k}   \right)^{2} } 
\det\limits_{1 \leqslant i,j \leqslant N}\psi_{i-1}\left(W+2\frac{x_{j}}{\sqrt{T}}\right) 
\det\limits_{1 \leqslant i,j \leqslant N}\psi_{i-1}^{*}\left( W S_{j}  \right) \ \left(\frac{T}{\tau_{0}}  \right) ^{- E_{\vec{k}_0}(N,W)}  \, .
\end{eqnarray}
  
We note that the eigenfunctions $\psi_i$ are defined on $[W, + \infty[$ such that $  \det\limits_{1 \leqslant i,j \leqslant N}\psi_{i-1}^{*}\left(W S_{j}  \right) $ is a well-defined constant. Injecting the equivalent expression for $Q(\vec{x},W)$ into Eq.~\eqref{eq:RasFunctionOfQ}, we see that the exponential terms cancel exactly such that $R(\vec{x},W) $ is equivalent to:
\bea
R(\vec{x},W) &\simeq&  \left[ \det\limits_{1 \leqslant i,j \leqslant N}\psi_{i-1}\left(W+2\frac{x_{j}}{\sqrt{T}}\right)  \det\limits_{1 \leqslant i,j \leqslant N}\psi_{i-1}^{*}\left(W S_{j}  \right)  \right]^2 \nn\\
& \times &
 \frac{ (4 \pi )^N \ P \left(\vec{A}(T) = \vec{0} \mid \vec{A}(\tau_0) = \vec{s} \right)^2}{P\left(   \vec{A}(T-\tau_0) = \vec{s} ,  \mathrm{B}^A_{[\tau_0,T-\tau_0]}  ,  \mathrm{NC}^A_{[\tau_0,T-\tau_0]}  \mid \vec{A}(\tau_0) = \vec{s} \right)} 
 \left(\frac{\tau_{0}}{T}  \right) ^{2 E_{\vec{k}_0}(N,W)}  \, .
 \eea
 The limit value of the propagator on the numerator is:
 \begin{equation}
  \lim_{\tau_0 \to 0} P \left(\vec{A}(T) =    \vec{0} \mid \vec{A}(\tau_0) = \vec{s} \right) 
  =\lim_{\tau_0 \to 0}  \bigg[ \frac{1}{\left( 2 \pi (T- \tau_0)  \right)^{N/2}} 
  e^{ - \frac{1}{2 (T - \tau_0) } \sum\limits_{k=1}^N s_k^2 }    \bigg]
  = \left( 2 \pi T \right)^{- N/2} \, .
 \end{equation}
 We finally obtain the following equivalent expression for $R(\vec{x},W)$ in the limit $\tau_0 \to 0$:
 \begin{equation}
 \label{eq:EquivalentExpressionRxw}
R\left(\vec{x},W\right) \simeq  \frac{\left(\frac{2}{T}\right)^{N}\left(\frac{\tau_{0}}{T}\right)^{2E_{\vec{k}_{0}}(N,W)}\left(\det\limits _{1\leqslant i,j\leqslant N}\psi_{i-1}^{*}\left(W S_{j}\right)\right)^{2}}{P\left(\vec{A}(T^{f})=\vec{s},\mathrm{B}_{[\tau_{0},T^{f}]}^{A},\mathrm{NC}_{[\tau_{0},T^{f}]}^{A}\mid\vec{A}(\tau_{0})=\vec{s}\right)} \left[\det\limits _{1\leqslant i,j\leqslant N}\psi_{i-1}\left(W+2\frac{x_{j}}{\sqrt{T}}\right)\right]^{2}  \, .
\end{equation}
In the goal to study this quantity as a probability distribution on the $\vec{x}$ variables, let us write the terms that do not depend on $\vec{x}$ in the r.h.s. of Eq. \eqref{eq:EquivalentExpressionRxw} as a prefactor $C_{N ,T }^{(\tau_0)}$. 
We thus write:
\begin{equation}
 R(\vec{x},W) \simeq
C_{N ,T }^{(\tau_0)}   \
\left[ \det\limits_{1 \leqslant i,j \leqslant N}\psi_{i-1}\left(W+2\frac{x_{j}}{\sqrt{T}}\right) \
  \right]^2 \, .
\end{equation}
We recall that the eigenfunctions $\psi_i$ are given explicitly in Eq. \eqref{eq:ExpressionPsiParabolicCylinder}. The normalization constant limit $C_{N ,T }=\lim_{\tau_0\to0} C_{N ,T }^{(\tau_0)}$ is fully determined by the following integral, which is readily computed by the continuous Cauchy-Binet, or Andr\'eief's, formula \cite{Andreief,Bruijn,ForresterMeetAndreief} and by the orthonormal properties of the 1-particle eigenfunctions $\psi$:
\begin{eqnarray}
C_{N ,T} ^{-1} &=& \int\limits_{0<x_1 < \cdots < x_N} \! \left[ \det\limits_{1 \leqslant i,j \leqslant N}\psi_{i-1}\left(W+2\frac{x_{j}}{\sqrt{T}}\right)   \right]^2  
   \prod_{k=1}^N \mathrm{d}x_k = \det\limits_{1 \leqslant i,j \leqslant N} \int\limits_{0}^\infty \psi_{i-1}\left(W+2\frac{x}{\sqrt{T}}\right)      \psi_{j-1}\left(W+2\frac{x}{\sqrt{T}}\right)      \mathrm{d}x \nn \\
&=& \det\limits_{1 \leqslant i,j \leqslant N} \delta_{i,j} \frac{\sqrt{T}}{2} = \frac{T^{N/2}}{2^{N}}\,.
\end{eqnarray}

We thus conclude that the limit $\{ \tau_0, \vec{s} \} \to \{ 0 , \vec{0}\}$ can be taken, as we do here, in a way that yields the following well-defined distribution on $\vec{x}$:
\begin{equation}
\label{eq:ResultRZeroInitialCondition}
\lim_{\tau_0 \to 0} R(\vec{x},W) =
C_{N ,T }   \
\left[ \det\limits_{1 \leqslant i,j \leqslant N}\psi_{i-1}\left(W+2\frac{x_{j}}{\sqrt{T}}\right) \
  \right]^2 \, .
\end{equation}
In the rest of this work, we assume that the small-$\tau_0$ limit of the distribution is taken and we denote it simply as $R(\vec{x},W)$. Writing $C_{N ,T }^{(\tau_0)}$ explicitly, we have obtained the following limit:
\begin{equation}
 \lim_{\tau_0\to 0}    \frac{   
 ( \frac{2}{T} )  ^{N} \left(\frac{\tau_{0}}{T}  \right) ^{2 E_{\vec{k}_0}(N,W) }   
 \left( \det\limits_{1 \leqslant i,j \leqslant N}\psi_{i-1}^{*}\left(W S_{j}  \right) \right)^2
 }{P\left(   \vec{A}(T-\tau_0) = \vec{s} ,  \mathrm{B}^A_{[\tau_0,T-\tau_0]}  ,  \mathrm{NC}^A_{[\tau_0,T-\tau_0]}  \mid \vec{A}(\tau_0) = \vec{s} \right)}
 = C_{N ,T} \, .
\end{equation}
The probability term in the denominator is then equivalent, as $\tau_0$ goes to zero, to:
\begin{equation}
P\left(   \vec{A}(T-\tau_0) = \vec{s} ,  \mathrm{B}^A_{[\tau_0,T-\tau_0]}  ,  \mathrm{NC}^A_{[\tau_0,T-\tau_0]}  \mid \vec{A}(\tau_0) = \vec{s} \right) \simeq 
 T^{- \frac{N}{2} } \left(\frac{\tau_{0}}{T}  \right) ^{2 E_{\vec{k}_0}(N,W) }   
 \left( \det\limits_{1 \leqslant i,j \leqslant N}\psi_{i-1}^{*}\left(W S_{j}  \right) \right)^2 \, .
\end{equation}
In particular we obtain the rate of the decay to zero as $\tau_0 \to 0$ with fixed $\vec{S}$: the probability to remain non-crossing and above the barrier decays with a factor of $\tau_0^{ E_{ \vec{k}_0}(N,W)   }$ for each one of the two problematic situations, close to $0$ and close to $T$. Furthermore, the influence of the fixed vector $\vec{S}$ is simply that the decay is in each case proportional to $\det\limits_{1 \leqslant i,j \leqslant N}\psi_{i-1}^{*}\left(W S_{j} \right)$, which also appears with a square.

 \subsection{Large $W$ limit}

As explained in section \ref{sec:mainResults}, we aim to probe the large $W$ limit where the wall pushes the particles in a large-deviation regime.
In this subsection, we obtain the behavior of the eigenfunctions $\psi$ in this regime. 

Let us evaluate the potential of Hamiltonian $\hat{H}$ at $X = W+\Delta X$ in the large $W$ limit. We note that $\Delta X \geqslant 0$ and $\Delta X =0$ is the position of the hard-wall, where all eigenfunctions vanish. The potential is:
\begin{eqnarray}
V(X) \ = \ \frac{1}{8} X^{2}-\frac{1}{4}
\ = \  \frac{1}{4} W \Delta X + \frac{1}{8} W^2 + \frac{1}{8} \Delta X^2 - \frac{1}{4} \, .
\end{eqnarray}
In the limit of large $W$, we neglect the quadratic term in $\Delta X$ such that the potential is the affine function:
\begin{equation}
\label{eq:Vapprox}
V(\Delta X) \simeq  \frac{1}{4} W \Delta X + \frac{1}{8} W^2 - \frac{1}{4} \, .
\end{equation}
Denoting $E_{\mathrm{offset}} = \frac{1}{8} W^2 - \frac{1}{4}$, the eigenfunctions $\tilde{\psi}(\Delta X) = \psi (W + \Delta X)$ solve the following Schr\"{o}dinger equation:
\begin{equation}
\hat{H}\tilde{\psi}=\left(\tilde{E}+E_{\mathrm{offset}}\right)\tilde{\psi}\quad\iff\quad\frac{\partial^{2}}{\partial\left(\Delta X\right)^{2}}\tilde{\psi}-\left(\frac{1}{2}W\;\Delta X\right)\tilde{\psi}=-2\tilde{E}\tilde{\psi}\,.
\end{equation}
Introducing a rescaled variable $Y$ such that $\Delta X=\left(\frac{2}{W}\right)^{1/3}Y+\frac{4\tilde{E}}{W}$ and $f(Y)=\tilde{\psi}\left(\left(\frac{2}{W}\right)^{1/3}Y+\frac{4\tilde{E}}{W}\right)$, we obtain the renowned Airy equation:
\begin{equation}
  \frac{\partial^{2}}{\partial Y^{2}} f - Y f = 0 
 \end{equation}
 with the following boundary condition stemming from the hard-wall condition at $X=W$:
\begin{equation}
f\left(-\left(\frac{2}{W}\right)^{2/3}2\tilde{E}\right)=0 .
\end{equation}
The solution that satisfies the boundary condition $f(Y\to\infty) \to 0$ is the Airy function $f(Y) \propto \operatorname{Ai}(Y)$, and the admissible values for $\tilde{E}$ are ${  \tilde{E}_k  = - \frac{\alpha_k}{2}   (\frac{W}{2})^{2/3} }$, with $k = 1,2,\dots$. We conclude that the $k$-th eigenfuntion $\psi_k$ and its corresponding energy $E_k$ for the Hamiltonian \eqref{Hamiltonian} is in the limit of large $W$ :
\begin{equation}
\label{eq:AiryEigenfunction}
\left\{ \begin{array}{ll}
\psi_{k}(X)=D_{k,W}\,\operatorname{Ai}\left(\alpha_{k}+\left(\frac{W}{2}\right)^{1/3}(X-W)\right)\\[5pt]
E_{k}=-\frac{\alpha_{k}}{2}(\frac{W}{2})^{2/3}+\frac{1}{8}W^{2}-\frac{1}{4}
\end{array}\right.
\end{equation}
where the normalization constant $D_{k,W} $ is such that $\int_W^\infty \psi_k^2 (X) dX = 1$. Using $\int_{\alpha_k}^\infty \operatorname{Ai}(x)^2 \mathrm{d}x = \operatorname{Ai}'(\alpha_k)^2$ one finds
\begin{equation}
D_{k,W}= \frac{(W/2)^{\frac{1}{6}}}{\abs{ \operatorname{Ai}^\prime ( \alpha_k ) }} \; .
\end{equation}

Let us now find the regime in which the large-$W$ approximations made in this subsection are valid. Neglecting the quadratic term in the potential as we did in Eq.~\eqref{eq:Vapprox} requires $\Delta X \ll W$. On the other hand, the eigenfunctions \eqref{eq:AiryEigenfunction} are spread over a spatial scale $\Delta X \sim W^{-1/3}$. We therefore find that the requirement is simply $W \gg 1$. Recalling that the wall's position is given by Eq.~\eqref{eq:gDefinition}, this requirement simply means that the distance reached by the wall, $g\left(T/2\right)\sim W\sqrt{T}$, is much larger than the typical diffusion length $\sim \sqrt{T}$ (we remind the reader that the diffusion constant is $D=1/2$), so that the wall ``pushes'' the particles into a large-deviation regime.

In the large $W$ regime, the desired $R(\vec{x},W)$ distribution yields, by injecting the  eigenfunctions \eqref{eq:AiryEigenfunction} in \eqref{eq:ResultRZeroInitialCondition}:
\begin{equation}
\label{eq:RNWithNormalization}
R(\vec{x},W)=B_{N,T,W}\ \left[\det\limits _{1\leqslant i,j\leqslant N}\operatorname{Ai}\left(\alpha_{i}+\frac{(4W)^{1/3}}{\sqrt{T}}x_{j}\right)\right]^{2} \mathbbm{1}_{0<x_1<\ldots<x_N} \;,
\end{equation}
where the constant
\begin{equation}
\label{eq:BNUW}
B_{N, T, W} =  C_{N ,T}  \ \prod_{k=1}^N D_{k,W}^2  =
\frac{ (4W)^{\frac{N}{3}}
}{
T^{\frac{N}{2}} \
\prod_{k=1}^N \operatorname{Ai}^\prime (\alpha_k)^2
}
\end{equation}
ensures its desired normalization
\begin{equation}
\int_{0\le x_{1}\le\cdots\le x_{N}<\infty}R\left(x_{1},\dots,x_{N},W\right)dx_{1}\cdots dx_{N}=1.
\end{equation} 
Plugging Eq.~\eqref{eq:BNUW} into \eqref{eq:RNWithNormalization}, one obtains Eq.~\eqref{eq:ResDet} from section \ref{sec:mainResults}.

\section{Determinantal structure and large-$N$ asymptotic behavior}

\label{sec:determinantalStructure}

\subsection{Determinantal structure}

We now exploit a mapping to trapped fermions to obtain the determinantal structure of the density and correlations of the particles' positions. 
One can interpret the joint probability density function (JPDF) \eqref{eq:RNWithNormalization} as the exact JPDF of the positions $\vec{x}$ of $N$ spinless noninteracting fermions (up to a factor $N!$ in the normalization of the distribution%
\footnote{
The difference between the two distributions is a multiplicative factor $N!$ that comes from the fact that for fermions, one doesn't require them to be ordered $x_{1}\le\cdots\le x_{N}$ (as we do require for non-crossing brownian bridges).
}). Choosing units so that the fermions' masses and $\hbar$ are equal to unity, these fermions are under the influence of the trapping potential 
\be
\label{eq:Vtildedef}
\tilde{V}\left(x\right)=\begin{cases}
\frac{2W}{T^{3/2}}x & x>0\\
\infty & x\le0
\end{cases} \; .
\ee
Then the normalized single-particle eigenfunctions of the Hamiltonian $\,-\frac{1}{2}\frac{d^{2}}{dx^{2}}+\tilde{V}\left(x\right)$ are 
given in Eq.~\eqref{eq:phidef},
and one indeed identifies that the determinant that appears in Eq.~\eqref{eq:RNWithNormalization} is a Slater determinant constructed from the $N$ lowest-energy eigenfunctions, $\phi_{1}\left(x\right),\dots,\phi_{N}\left(x\right)$. Note that the $\phi$ eigenfunctions are simply obtained from the large-W $\psi$ functions of the previous section through the change of variables $X \to W + \frac{2x}{\sqrt{T}}$.
Therefore, we find that the JPDF \eqref{eq:RNWithNormalization} can be written as a \emph{single} determinant, see Eq.~\eqref{eq:RNSingleDeterminant},
with the kernel $K_{N}\left(x,x'\right)$ given in Eq.~\eqref{eq:kernel_def}.
Thus, the JPDF \eqref{eq:RNWithNormalization} describes a determinantal point process \citep{Borodin1999}. Such processes have been studied extensively and many general results have been obtained for them, which we  use below.

\begin{figure}[h!]
  \includegraphics[width=0.6\linewidth]{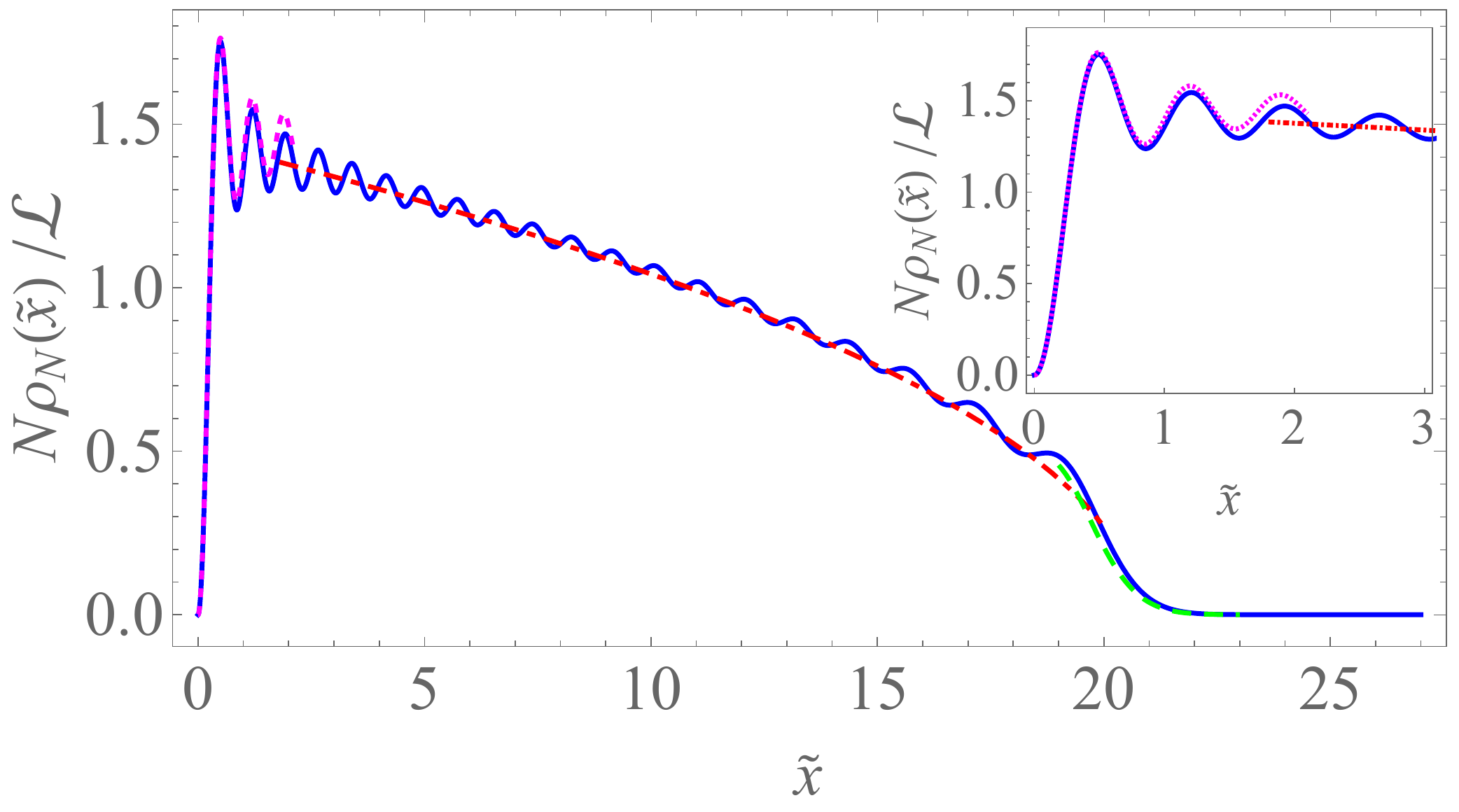}
  \caption{Plot of the rescaled density $N \rho_N / \mathcal{L}$ versus the rescaled distance $\tilde{x}=x/\mathcal{L}$ from the wall, where $\mathcal{L}=\sqrt{T}/\left(4W\right)^{1/3}$, for $N=20$. The solid blue line is a direct evaluation of $N\rho_{N}\left(x\right)=K_{N}\left(x,x\right)$ using the expression for the kernel as a sum over eigenfunctions \eqref{eq:kernel_def}. 
The dot-dashed red line, dashed green line and dotted magenta line are the approximate expressions in the bulk regime  \eqref{eq:bulk_density}, near the soft edge \eqref{eq:densitySoftEdge}, and near the wall \eqref{eq:densityHardEdge}, respectively. The inset is a zoom-in on the region near the wall.}
  \label{fig:density}
\end{figure}

Importantly the kernel~\eqref{eq:kernel_def} satisfies the ``reproducibility'' property
\be
\label{eq:reproducibility}
\int_{0}^{\infty}K_{N}\left(x,z\right)K_{N}\left(z,y\right)dz=K_{N}\left(x,y\right).
\ee
It is straightforward to obtain Eq.~\eqref{eq:reproducibility} by using the orthonormalization of the eigenfunctions $\int\phi_{k}^{*}\left(x\right)\phi_{k'}\left(x\right)=\delta_{k,k'}$.
Equivalently, the kernel is a projector: $\hat{K}_{N}^{2}=\hat{K}_{N}$ in operator notation, as follows immediately from its definition $\hat{K}_{N}=\sum_{k=1}^{N}\left|\phi_{k}\right\rangle \left\langle \phi_{k}\right|$. Eq.~\eqref{eq:kernel_def} is the representation of $\hat{K}_{N}$ in the position basis: $K_{N}\left(x,y\right)=\left\langle x\,|K_{N}|\,y\right\rangle $. 
A general property of determinantal point processes \cite{Forrester, Mehta} is that any $n$-point correlation function, where $1 \le n \le N$, can be written as an $n\times n$ determinant of a matrix whose entries are given by the kernel $K_N$.  
For the particular case $n=1$, one finds that
the (average) density of particles is given by
\begin{equation}
\label{eq:density_from_kernel}
N\rho_{N}\left(x\right)=\left\langle \sum_{i=1}^{N}\delta\left(x-x_{i}\right)\right\rangle =K_{N}\left(x,x\right),
\end{equation}
where here (and below) the averaging is defined is over the JPDF $R(\vec{x},W)$:
\begin{equation}
\left\langle *\right\rangle \equiv\int_{0\le x_{1}\le\cdots\le x_{N} < \infty}\left(*\right)R\left(x_{1},\dots,x_{N},W\right)dx_{1}\cdots dx_{N}.
\end{equation}
Note that, in our conventions, the density is normalized to unity (and not to the number of particles) $\int_{0}^{\infty}\rho_{N}\left(x\right)dx=1$.
In Fig.~\ref{fig:density} we plot the density $N\rho_{N}\left(x\right)$ for $N=20$ particles.
It is worth mentioning that other observables, such as the full counting statistics of the number of particles in a given interval, as well as the distribution of the position of the ``top'' particle $x_N$ and of the ``bottom'' particle $x_1$, can be extracted from the kernel by calculating Fredholm determinants, see for example Ref.~\cite{DeanLeDoussal16} for details.

\subsection{Large-$N$ limit}

In the limit where the number of Brownian bridges is very large, $N\gg1$, many of the important observables described above approach universal behaviors which are well known from the theory of noninteracting fermions and/or from random matrix theory (RMT) \cite{DeanLeDoussal16,DeanLeDoussal19}. These behaviors are different in (i) the bulk regime, which describes the range of distances from the wall where most of the particles typically are, (ii) the soft-edge regime, which describes the vicinity of the typical position of the ``top'' particle $x_N$, and (iii) the hard-wall edge regime, which describes very short distances from the wall, of the order of the typical position of the ``bottom'' particle $x_1$. In this section we briefly describe the universal behavior in each of these three regimes, and we refer the reader to the recent review \cite{DeanLeDoussal19} for further details of the derivations of these results.

\subsubsection{Bulk regime}

In the bulk regime, the density is well described \cite{DeanLeDoussal16} by the local density approximation (LDA) \cite{Castin06}. This approximation gives $\rho_{N}\left(x\right)\simeq\rho^{\text{bulk}}\left(x\right)$ with
\begin{equation}
\label{eq:bulk_density_general}
N\rho^{\text{bulk}}\left(x\right)=\frac{\sqrt{2}}{\pi}\left[\mu-\tilde{V}\left(x\right)\right]_{+}^{1/2}\Theta\left(x\right),\qquad \left(z\right)_{+}^{1/2}\equiv \begin{cases}
z^{1/2} & z>0\\
0 & z<0
\end{cases}\,,
\end{equation}
where $\Theta\left(x\right)$ is the Heaviside function, and $\mu$ is the (effective) Fermi energy of the corresponding fermion system, and is related to the total number of particles
$N$ via
\begin{equation}
\label{eq:NasFunctionOfMu}
N=\int_{0}^{\infty}N\rho_{N}\left(x\right)dx\simeq\int_{0}^{\infty}dx\,\frac{\sqrt{2}}{\pi}\left[\mu-\tilde{V}\left(x\right)\right]_{+}^{1/2}.
\end{equation}
Plugging our potential \eqref{eq:Vtildedef}  into Eq.~\eqref{eq:NasFunctionOfMu}, we obtain $N\simeq\sqrt{2} \, T^{3/2}\mu^{3/2}\!/3\pi W$, or $\mu=\left(3\pi WN\right)^{2/3}\!/2^{1/3}T$, and then Eq.~\eqref{eq:bulk_density_general} becomes
\be
\label{eq:bulk_density}
N\rho^{\text{bulk}}\left(x\right)=\frac{\sqrt{2}}{\pi}\left[\frac{\left(3\pi WN\right)^{2/3}}{2^{1/3}T}-\frac{2W}{T^{3/2}}x\right]_{+}^{1/2}\Theta(x),
\ee
see Fig.~\ref{fig:density}.
In the bulk, the kernel \eqref{eq:kernel_def}  
converges to the celebrated sine kernel \cite{Kamien1988}
\be
\label{eq:sine_kernel_def}
K_N\left(x,y\right)\simeq\frac{\sin\left[k_{F}\left(x\right)\left(x-y\right)\right]}{\pi\left(x-y\right)}
\ee 
where $k_{F}\left(x\right)=\sqrt{2\left[\mu-\tilde{V}\left(x\right)\right]}=\pi N\rho^{{\rm bulk}}\left(x\right)$ is the local Fermi wave vector of the corresponding fermionic system.

\subsubsection{Soft edge regime}

The LDA breaks down near the edge which is defined by $\tilde{V}\left(x_{\text{edge}}\right)=\mu$, which here gives 
\be
x_{\text{edge}}=\frac{\left(3\pi N\right)^{2/3}T^{1/2}}{2^{4/3}W^{1/3}}.
\ee
Near the edge, the density is correctly described by \cite{DeanLeDoussal16}
\begin{equation}
\label{eq:densitySoftEdge}
N\rho_{N}\left(x\right)\simeq\frac{1}{w_{N}}F_{1}\left(\frac{x-x_{\text{edge}}}{w_{N}}\right),\qquad
 F_{1}\left(z\right)=\left[\text{Ai}'\left(z\right)\right]^{2}-z\left[\text{Ai}\left(z\right)\right]^{2}
\end{equation}
where $w_{N}=\left[2\tilde{V}'\left(x_{\text{edge}}\right)\right]^{-1/3} \! =T^{1/2}/\left(4W\right)^{1/3}$ is the width of the edge regime, see Fig.~\ref{fig:density}.
 Furthermore, in the edge regime, the kernel \eqref{eq:kernel_def} converges to
\be
\label{softkernel} 
K_N\left(x,y\right)\simeq\frac{1}{w_{N}}K_{{\rm Ai}}\left(\frac{x-x_{\text{edge}}}{w_{N}},\frac{y-x_{\text{edge}}}{w_{N}}\right)
\ee
in terms of the Airy kernel \citep{Bowick91, TracyWidom94}
\be
\label{airy_kernel.1}
K_{\rm Ai}(a,b) =
\frac{{\rm Ai}(a){\rm Ai}'(b)-{\rm Ai}'(a){\rm Ai}(b)}{a-b} 
=  \int_0^{+\infty} du \, {\rm Ai}(a+u) {\rm Ai}(b+u) \, .
\ee 
As a result, the fluctuations of the position of the ``top'' particle (furthest from the wall) behave as $x_{N}=x_{\text{edge}}+w_{N}\chi$ where $\chi$ converges (in the $N \gg 1$ limit) to the celebrated Tracy-Widom Gaussian unitary ensemble (GUE) distribution \cite{TracyWidom94}.

\subsubsection{Hard-wall edge regime $x \simeq 0$}

The LDA also breaks down very close to the wall. Here the kernel \eqref{eq:kernel_def} converges to 
\be
K_{N}\left(x,y\right)\simeq k_{F}K_{{\rm Hb}}\left(k_{F}x,k_{F}y\right)
\ee
where $k_{F}=k_{F}\left(x=0\right)=\sqrt{2\mu}=\left(6\pi WN\right)^{1/3} \! /\sqrt{T}$ and 
\be
\label{eq:KHbdef}
K_{{\rm Hb}}\left(z,z'\right)=\frac{\sin\left(z-z'\right)}{\pi\left(z-z'\right)}-\frac{\sin\left(z+z'\right)}{\pi\left(z+z'\right)}
\ee
is the universal hard-wall kernel (a particular case of the Bessel kernel) \cite{TracyWidomBessel94, TracyWidom07, Calabrese11,LLMS17,LLMS18,DeanLeDoussal19,Gautie21}.
By taking the limit $x \to y$, we obtain the density near the wall
\be
\label{eq:densityHardEdge}
N\rho_{N}\left(x\right) = K_{N}\left(x,x\right) \simeq k_{F}F_{{\rm Hb}}\left(k_{F}x\right),\qquad F_{{\rm Hb}}\left(z\right)=\frac{1}{\pi}\left[1-\frac{\sin\left(2z\right)}{2z}\right],
\ee
see the inset of Fig.~\ref{fig:density}.
The PDF $P\left(x_{1}\right)$ of the position of the ``bottom particle'' $x_1$ (the particle closest to the wall), in the typical fluctuation regime, takes the scaling form $P\left(x_{1}\right)\simeq k_{F}\mathcal{P}_{\text{Hb}}\left(k_{F}x_{1}\right)$ where $\mathcal{P}_{\text{Hb}}\left(z\right)$ is a universal scaling function that can be expressed in terms of a Fredholm determinant that involves the hard-wall kernel \eqref{eq:KHbdef}, see Ref.~\cite{LLMS17} for details.

\section{Conclusion}  
\label{sec:conclusion}

In this paper, we considered $N$ non-crossing Brownian bridges conditioned to avoid a wall whose position as a function of time $g(\tau)$ is given by a semicircle, thereby extending the Ferrari-Spohn model which corresponds to $N=1$. We calculated the joint distribution $R\left(\vec{x},W\right)$ of the distances between the Brownian particles and the wall at an intermediate time. We achieved this by mapping this problem to that of $N$ non-crossing Brownian motions under a moving wall whose position is proportional to the square root of time. Focusing on the limit where the wall $g(\tau)$ pushes the Brownian bridges into a large-deviation regime, we showed that the joint distribution $R\left(\vec{x},W\right)$ coincides with that of $N$ noninteracting spinless fermions trapped by a linear potential and a hard wall. We exploited the connection to fermions in order to uncover universal behavior of the density of particles and the density-density correlations that emerge when the number of particles is large, $N\gg1$. In particular, this enabled us to calculate the distributions of the positions of the top and bottom particles, and we found that the former is described by the Tracy-Widom GUE distribution.

Here we considered only the particular case of the semicircle wall \eqref{eq:gDefinition} with the observation time $\tau_{\text{obs}}=T/2$, and we assumed that the diffusion constant was $D=1/2$. However, it is natural to expect our results to hold for any moving wall described by a function $g\left(\tau\right)$ that satisfies $g\left(0\right)=g\left(T\right)=0$ and $g''\left(\tau\right)<0$, for any observation time (and for any diffusion constant), provided the wall ``pushes'' the particles into a large-deviation regime. The factor $\left(4W\right)^{1/3}\!/\sqrt{T}$ that appears in our results, e.g., in \eqref{eq:ResDet} and
\eqref{eq:phidef}, should be replaced by $\left[-g''\left(\tau_{\text{obs}}\right)/2D^{2}\right]^{1/3}$. This general result is known to be correct for $N=1$ \citep{FerrariSpohn05}.
For such functions $g(\tau)$, one could still use the mapping \eqref{eq:AToBMapping} that we used in this work. However, the wall barrier $h(t)$ would not be the square root barrier $W\sqrt{t}$, so the results of Ref. \cite{NonCrossing} would not be directly applicable.
A further interesting extension would be to consider the multi-time distribution: the joint distribution of the positions of the particles at observation times $\tau_{\text{obs},1},\dots,\tau_{\text{obs},N}$. Multi-time results, for other models, were obtained in \cite{Prahofer2002,Ferrari2008}.

Our results are valid for typical fluctuations of the distances of the particles from the wall, and it would be interesting to study large deviations as well. This was done for the particular case $N=1$ in \cite{GeomOpt1} (further large-deviation properties in the Ferrari-Spohn model were studied in \cite{GeomOpt2}).
We expect the results of \cite{GeomOpt1} to hold for arbitrary $N$ for large deviations of the position of the top particle, because the effect of the other $N-1$ particles will be negligible in the leading order, as they will typically stay relatively close to the wall.
Finally, the Ferrari-Spohn distribution was encountered in \cite{Agranov19} in the context of a single Brownian excursion conditioned to cover a small area $\int x(t) dt$. It is natural to expect our results to be valid for $N$ non-crossing Brownian excursions $x_i(t)$ conditioned on the sum of areas $\sum_{i=1}^{N}\int x_{i}(t)dt$ being small. Similar models, of non-crossing Brownian particles near a hard wall subject to area tilts, were recently studied in Refs.~\cite{Ioffe2018a, Ioffe2018c},
and an expression similar to our Eq.~\eqref{eq:ResDet} was obtained in \cite{Ioffe2018a}, supporting this conjecture.

\acknowledgements

We thank Pierre Le Doussal for useful discussions and for suggesting the problem studied here, and Gregory Schehr and Satya N. Majumdar for useful discussions. NRS acknowledges support from the Yad Hanadiv fund (Rothschild fellowship) and ANR grant ANR-17-CE30- 0027-01 RaMaTraF.

 \end{document}